\documentstyle[psfig]{article}
\input amssym.def
\newcommand{\eq}{\begin{equation}}
\newcommand{\eqx}{\end{equation}}
\newcommand{\eqn}{\begin{eqnarray}}
\newcommand{\eqnx}{\end{eqnarray}}
\newcommand{\psiz}{\Psi_{z_0}(z_1,z_2,z_3)}
\newcommand{\dr}{\partial}
\newcommand{\f}[2]{\frac{#1}{#2}}
\newcommand{\Z}{{\Bbb{Z}}}
\newcommand{\C}{{\Bbb{C}}}
\newcommand{\slc}{$SL(2,\C)$ }
\newcommand{\slz}{$SL(2,\Z)$ }
\newcommand{\lm}{\lambda}
\newcommand{\om}{\omega}
\newcommand{\ra}{\longrightarrow}
\newcommand{\HH}{{\cal{H}}}
\newcommand{\dd}[2]{\f{d^{#1}}{d#2^{#1}}}
\newcommand{\br}[1]{\bar{#1}}

\newcommand{\rr}[4]{#1, { #2 \/}{ #3} #4}

\title{The Odderon and Invariants of Elliptic Curves.\footnote{
Supported in part
by KBN grant no. 2 P03B 083 08}\\
{ \normalsize Talk presented at the Cracow Epiphany Conference on Proton
Structure, Krakow, Poland, January 5-6, 1996}}
\author{Romuald A. Janik\\
\small Institute of Physics, Jagellonian University, Cracow\\
\small Reymonta 4, 30-059 Krakow\\
\small e-mail: ufrjanik@jetta.if.uj.edu.pl}

\date{}

\begin{document}

\maketitle

\begin{abstract}
In this talk we present some links of the theory of the odderon with
elliptic curves.
These results were obtained in an earlier work \cite{RJ}. 
The natural degrees of freedom of the odderon turn
out to coincide with conformal invariants of elliptic curves with a
fixed `sign'. This leads to a formulation of the odderon which is
modular invariant with respect to $\Gamma^2$ --- the
unique normal subgroup of \slz {\,} of index 2.
\end{abstract}

\vfill\noindent{TPJU-9/96\\}
\pagebreak

\section{Introduction}

Recently the study of the small Bjorken $x$ region of Deep Inelastic
Scattering has attracted much attention. The behavior of the
structure functions in this limit turns out to be governed by the
exchange of Regge poles. The one most important for the $F_2$
structure function is the $C=+1$ pole with vacuum quantum numbers ---
the BFKL pomeron, described in the framework of perturbative QCD as
the exchange of two `reggeized' gluons.
Already in the late 70's 
this amplitude was calculated,
and the formula for the intercept of the
renowned BFKL pomeron was derived (see \cite{BFKL} \cite{Lip}). 
Lipatov's solution depends in a crucial way 
on the global conformal symmetry of the problem. 
The next step led to the derivation of the
Bartels-Kwieci{\'n}ski-Prasza{\l}owicz equation describing the exchange
of three reggeized gluons (\cite{KwPrasz},\cite{GLLA}). The
corresponding $C=-1$ pole, called the odderon is thought to have
influence on the $F_3$ structure function. Later this approach was
extended to the case of an arbitrary number of reggeons
in the form of the Generalized Leading Logarithm Approximation (GLLA)
\cite{GLLA}. 
It is important to emphasize that the odderon should not be treated
just as a correction to the pomeron, but that it has it's own
distinct physical signature.

The theory of the odderon turned out to be much more difficult to
solve than the BFKL pomeron case, and a wide variety of approaches
have been attempted. Among them is the very surprising connection with
exactly solvable lattice models, namely the Heisenberg XXX s=0 spin
chain (\cite{LipSolv1}, \cite{padwa}, \cite{KorFad}, \cite{Kor}).
Within this framework variants of the Bethe ansatz have been tried
(\cite{KorFad}, \cite{LipSolv2}, \cite{Kor},
\cite{SW}), quasiclassical approximation\cite{QCl}, but still the
explicit value of the odderon intercept is unknown, apart from some
variational bounds (\cite{GauLip}).

In this talk we will present the results obtained earlier in
\cite{RJ}, that the theory of the odderon possesses modular
invariance, well known from conformal field theory and string theory.
This observation follows from an intriguing link of the odderon with
invariants of elliptic curves.
First we will recall the theory of the odderon, then, following Lipatov,  
the consequences of global \slc
invariance.  The new results in \cite{RJ} were the analysis of the
role of cyclic symmetry in this framework and the link with modular
invariance through the theory of elliptic curves.

\section{The Odderon}

The Regge limit of QCD is defined as the kinematical region
\eq
s\gg -t \approx M^2
\eqx
where $M$ is the hadron mass scale, or,  in the case of Deep Inelastic
Scattering, as the small $x=Q^2/s$ limit. Here we sketch
the equivalence between the Regge intercept of 
amplitudes and energy levels of a two-body  Hamiltonian within the
GLLA approximation.

%\psdraft
\begin{figure}[t]
\vspace{1cm}
\psfig{file=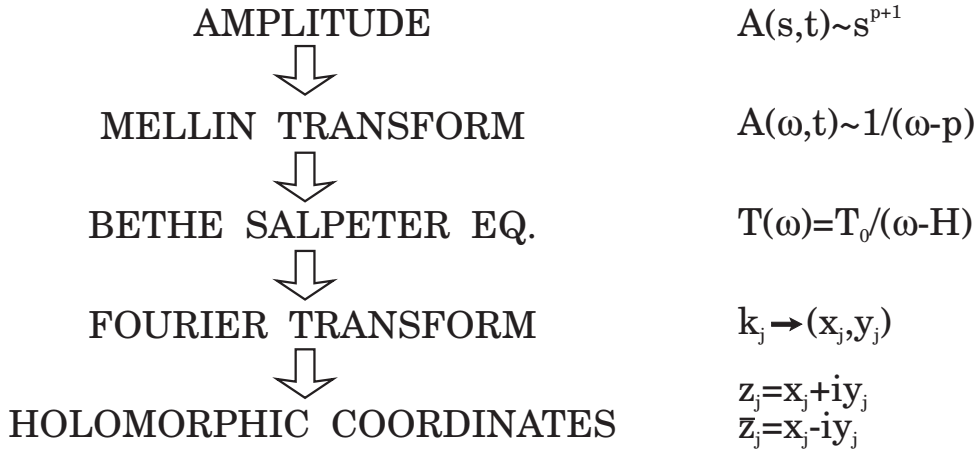,height=6cm,bburx=0cm}
\vspace{1cm}
\label{f.steps}
\caption{The main steps leading to the equivalence of the Regge 
intercept and energy levels of a certain hamiltonian.}
\end{figure}

The main steps leading to this equivalence are briefly summarized in
figure 1.
The aim is to find the Regge behavior of the amplitude $A(s,t)\sim
s^{\om_0+1}$. 
This amplitude is described in the framework of perturbative QCD as a
sum of graphs corresponding to the exchange of $N$ reggeized gluons
with all possible gluonic interactions between them. The BFKL pomeron
corresponds to graphs with $N=2$, while the odderon is the $N=3$
contribution ( apart from higher $N$ corrections ). The Regge behavior of the
amplitudes can be directly translated, in the case of DIS, into the
$x$-dependence of the appropriate proton structure functions at low
$x$.

In the first step the power behavior  of the amplitude
is translated into singularities of the Mellin transform
\eq
A(s,t)=is\int_{\delta-i\infty}^{\delta+i\infty}\f{d\om}{2\pi i}
\left(\f{s}{M^2}\right)^\om A(\om,t)
\eqx
Rewriting this amplitude as the convolution of ``hadron''  wave
functions $\Phi_{A,B}$ and a kernel $T(\{k_i\},\{k'_j\},\om)$ we get:
\eq
A(\om,t)=\int d^2k_i \int d^2k'_j \Phi_A(\{k_i\})T(\{k_i\},\{k'_j\},\om)
\Phi_B(\{k'_j\})
\eqx
where $\{k_i\}$ and $\{k'_j\}$ are the transverse momenta of the $N$
exchanged reggeons ( in the case of odderon $N=3$ ).
The rest of this work deals with the singularity structure of the
kernel $T(\{k_i\},\{k'_j\},\om)$.
The next step amounts to writing the Bethe-Salpeter equations for the
kernel $T$:
\eq
\om T(\om)=T_0+\HH T(\om)
\eqx
which corresponds to the iteration of gluon interactions between the
reggeons.
Here $T_0$  is the free propagator and $\HH$ is the 
operator corresponding to the insertion of single gluonic interactions 
between all pairs of reggeons. This equation can be formally solved:
\eq
T(\om)=\f{T_0}{\om-\HH}
\eqx
It is clear now that in order to find the Regge intercept  it
suffices to find the eigenvalues of the Hamiltonian operator $\HH$.
The last step is to go to transverse `impact parameter' space and
introduce holomorphic coordinates.
After performing   Fourier transformation
( $k_i\ra b_i$) and  using the complex notation $z_j:=x_j+i
y_j$, the Hamiltonian splits into a sum of a holomorphic part and an
antiholomorphic part. In the large $N_c$ limit the
two commute. Due to this so-called holomorphic separability we may
seek eigenfunctions of the Hamiltonian as a product of eigenfunctions
of the holomorphic and antiholomorphic operators. Before we discuss
the odderon case let us recall the description of the BFKL pomeron in
this framework.

\subsection{The BFKL Pomeron}

The BFKL pomeron corresponds to the exchange of 2 reggeons. The
holomorphic and antiholomorphic hamiltonians are given by:
\eqn
\label{e.hbfkl}
H(z_1,z_2)&=&\sum_{l=0}^{\infty}\f{2l+1}{l(l+1)-L^2_{12}}-\f{2}{l+1}\\
H(\br{z}_1,\br{z}_2) &=&
\sum_{l=0}^{\infty}\f{2l+1}{l(l+1)-\br{L}^2_{12}}-\f{2}{l+1}
\eqnx
where
\eqn
L^2_{12} &:=& -z_{12}^2\f{d}{dz_1}\f{d}{dz_2}\\
\br{L}^2_{12} &:=& -\f{d}{d\br{z}_1} \f{d}{d\br{z}_2}\br{z}_{12}^2
\eqnx
being the holomorphic and antiholomorphic Casimir operators of the 
group \slc. Although the problem seems at first glance to be quite
intractable, it is in fact quite easy to solve. The crucial
ingredient is the \slc  invariance of the system. This enables us to
consider wavefunctions in a definite unitary representation of \slc,
and simply insert the eigenvalues of the Casimir operators into 
(\ref{e.hbfkl}) to obtain the energy.

The celebrated BFKL solution ($N=2$ case) corresponds in this language to 
finding the maximal eigenvalue of the equations
\eq
H(z_1,z_2)\Psi(z_1,z_2)=E\Psi(z_1,z_2) \quad
\br{H}(\br{z}_1,\br{z}_2)\Psi(\br{z}_1,\br{z}_2)=E\Psi(\br{z}_1,\br{z}_2)
\eqx
and has the known solution
\eq
E=-[\psi(m)+\psi(1-m)-2\psi(1)]\quad \br{E}=-[\psi(\br{m})
+\psi(1-\br{m})-2\psi(1)]
\label{ccc}
\eqx
where $\psi$ is the derivative of the logarithm of the  Euler $\Gamma$ function
and $m$ is a conformal weight. The maximum of (\ref{ccc}) is achieved at 
$m=1/2$ and reproduces the BFKL slope
\eq
\omega_0^{BFKL}=\frac{2\alpha_s N_c}{4\pi}(E+\br{E})=
\frac{\alpha_s N_c}{\pi}4\ln 2
\eqx

\subsection{The Odderon}

In the case of the odderon the problem looks deceptively similar. Now
the ( holomorphic ) hamiltonian is given by the sum of three terms, each
identical to the ordinary BFKL hamiltonian, namely:
\eq
(H(z_1,z_2)+H(z_2,z_3)+H(z_3,z_1))\Psi(z_1,z_2,z_3)=E\Psi(z_1,z_2,z_3)
\label{e.hamilt}
\eqx
where $H(z_i,z_j)$ is given by the same expression as (\ref{e.hbfkl}). 
The eigenvalue $E$ 
of the holomorphic hamiltonian and the corresponding eigenvalue $\bar{E}$
of the antiholomorphic one are related to the Regge intercept by the formula:
\eq
\om_0=\f{\alpha_s N_c}{4\pi}(E+\bar{E})
\eqx

The reason why this makes the problem at least an order of magnitude
more difficult is the fact that the three terms in (\ref{e.hamilt}) do not
commute. Furthermore there is no natural small parameter in which one
might try to perform a kind of perturbative expansion.

\subsection{\slc  invariance}

\label{s.lipstrat}

Since the global \slc invariance has proved to be so powerful as to
solve the BFKL pomeron, it is natural to try to use it to simplify
the problem also in the case of the odderon. This analysis has been
done by Lipatov \cite{LipAns}.

The Hamiltonian $H$ is invariant
with respect to the action of \slc 
on holomorphic functions given by:
\eq
(g\cdot \Psi)(z_1,z_2,z_3)=\Psi\left(\f{az_1+b}{cz_1+d},
\f{az_2+b}{cz_2+d},\f{az_3+b}{cz_3+d}\right)
\quad \hbox{ for }g=
\left(\begin{array}{cc} 
a&c\\ 
b&d
\end{array}\right) \in\hbox{\slc} 
\eqx 
Therefore it commutes with the holomorphic
Casimir operator for this representation:
\eq
\hat{q}_2:=
-z_{12}^2\f{d}{dz_1}\f{d}{dz_2}
-z_{23}^2\f{d}{dz_2}\f{d}{dz_3}
-z_{31}^2\f{d}{dz_3}\f{d}{dz_1}
\eqx
where $z_{ij}=z_i-z_j$.
This enables us to consider functions transforming under the unitary 
representations of \slc {\,} 
labelled by $n\in\Bbb{N}$ and $\nu\in\Bbb{R}$. In this case
the eigenvalue $q_2$ is $((1+n)/2+i\nu)((-1+n)/2+i\nu)$. 

Lipatov \cite{LipAns} has chosen an ansatz, which automatically diagonalizes 
$\hat{q}_2$:
\eq
\psiz={\left(\f{z_{12}z_{23}z_{31}}{z_{10}^2z_{20}^2z_{30}^2}
\right)}^{m/3}\varphi(\lm)
\eqx
where $m=1/2+i\nu+n/2$, $n$ is an integer and $\nu$ is a real number. 
Here, $z_0\in \C$ is just a parameter and
$\lm$ is the anharmonic ratio:
\eq
\lm=\f{z_{12}z_{30}}{z_{13}z_{20}}
\eqx

%This ansatz was also motivated by a conjectural link of the Odderon
%and Conformal Field Theory.

A breakthrough occurred when Lipatov \cite{Lipq3} established the 
existence of another integral of motion --- an operator $\hat{q}_3$:
\eq
\hat{q}_3=z_{12}z_{23}z_{31}\dr_1\dr_2\dr_3
\eqx
which commutes with the hamiltonian $H$.

Lipatov further derived the form of the operator $\hat{q}_3$ within
this ansatz. Inserting $\psiz$ into the equation
\eq
\label{e.cons}
\hat{q}_3\psiz=q_3\cdot\psiz
\eqx
and canceling the factor $(\ldots)^{m/3}$ he obtained:
\eq
\label{e.q3lip}
\nabla_1\f{1}{\lm(1-\lm)}\nabla_2\nabla_3\varphi(\lm)=q_3 \varphi(\lm)
\eqx
where
\eqn
\nabla_1 &=& \f{m}{3}(1-2\lm)+\lm(1-\lm)\dr,\\
\nabla_2 &=& \f{m}{3}(1+\lm)+\lm(1-\lm)\dr,\\
\nabla_3 &=& -\f{m}{3}(2-\lm)+\lm(1-\lm)\dr
\eqnx

One of the strategies  for solving the odderon problem,
proposed by Lipatov 
\cite{Lipq3}, was to diagonalize the conservation laws $\hat{q}_2$ and
$\hat{q}_3$ and to substitute the solution into the Schroedinger equation
in order to find the energy eigenvalue.

Unfortunately no one has succeeded in doing this. So it is quite
natural to seek for a new symmetry which might be powerful 
enough to obtain some progress.

%The Hamiltonian (\ref{e.hamilt}) has also been rewritten in terms of $\lm$. 

\section{Cyclic invariance}

It is easy to see that both the Hamiltonian $H$ and $\hat{q}_3$ are
invariant under cyclic permutations of the gluonic coordinates
$z_1,z_2,z_3$. We  show now how this symmetry manifests itself in the formalism
of the preceding section.
Under the permutation $z_1\ra z_2\ra z_3$
the anharmonic ratio transforms as follows:
\eq
\label{e.transf}
\lm \ra 1-\f{1}{\lm} \ra \f{1}{1-\lm}
\eqx
As we are interested mainly in obtaining the leading behavior of the
odderon amplitudes, which corresponds to finding the energy of the
ground state of the system, it is natural to postulate that the
relevant eigenfunction is symmetric under this
transformation and  so 
\eq
\varphi(\lm)=f(s_1,s_2,s_3,\tilde{j})
\eqx
where $s_i$ are the symmetric polynomials in $x_1=\lm, x_2=1-1/\lm$ and
$x_3=1/(1-\lm)$, and $\tilde{j}$ is the Vandermonde determinant.
Namely
\eqn
s_1 &=& x_1+x_2+x_3=\f{\lm^3-3\lm+1}{\lm(\lm-1)} \\
s_2 &=& x_1x_2+x_2x_3+x_3x_1=\f{\lm^3-3\lm^2+1}{\lm(\lm-1)} \\
s_3 &=& x_1x_2x_3= -1\\
\tilde{j} &=& (x_1-x_2)(x_2-x_3)(x_3-x_1) = \f{(\lm^2-\lm+1)^3}{\lm^2
(\lm-1)^2}
\eqnx
It turns out that the only independent quantity is
$A=s_1+s_2=\f{(\lm+1)(2\lm-1)(\lm-2)}{\lm(\lm-1)}$ related to $\tilde{j}$ by
the equation $4\tilde{j}=A^2+27$.
It is convenient to introduce the notation:
\eqn
B&:=&8A=\sqrt{j-1728} \label{e.sqrt}\\
j&:=&256 \tilde{j}=2^8  \f{(\lm^2-\lm+1)^3}{\lm^2(\lm-1)^2}
\label{e.vander}
\eqnx

At this moment we make a refinement of Lipatov's ansatz, namely

\eqn
\psiz&=&{\left(\f{z_{12}z_{23}z_{31}}{z_{10}^2z_{20}^2z_{30}^2}
\right)}^{m/3}f(B)\nonumber \\
&=&
{\left(\f{z_{12}z_{23}z_{31}}{z_{10}^2z_{20}^2z_{30}^2}
\right)}^{m/3}f\left(8\f{(\lm+1)(2\lm-1)(\lm-2)}{\lm(\lm-1)}
\right)
\eqnx
where $m=1/2+i\nu+n/2$, $n$ is an integer and $\nu$ is a real number. 
$\lm$ is the anharmonic ratio:
\eq
\lm=\f{z_{12}z_{30}}{z_{13}z_{20}}
\eqx

Now we insert the function $\varphi(\lm)=f(B)$ into the conservation law
(\ref{e.q3lip}). After reexpressing the result in terms of $j$ and 
$B=\sqrt{j-1728}$  we get:

\eq
\label{e.q3ost}
\Biggr\{\f{j^2}{2}\dd{3}{B} +2Bj\dd{2}{B}+ (j(1+\f{m(1-m)}{6}
)-3\cdot 2^8)\f{d}{dB}+
\f{(m-3)m^2}{27}B-8q_3\Biggl\}f(B)=0
\eqx

The advantage of considering this equation is that all the discrete
symmetries present in the form of the nonlinear transformation
(\ref{e.transf}) act trivially on this equation. The variable $j$  ( or
really $\sqrt{j-1728}$ ) seems to be the true physical variable of
the theory. There are no additional residual symmetries which one
could take into account. In the next section we give a geometrical
interpretation of the $j$ variable in terms of elliptic curves. This
will enable us to rephrase the theory of the odderon in a modular
invariant way.

%The original Hamiltonian (\ref{e.hamilt}) expressed by Lipatov in terms of
%$\lm$ can also be recast using the functions $B=\sqrt{j-1728}$ (although
%obtaining an explicit expression seems to be highly non-trivial). 

\section{Modular invariance}

\begin{figure}
\vspace{1cm}
\psfig{file=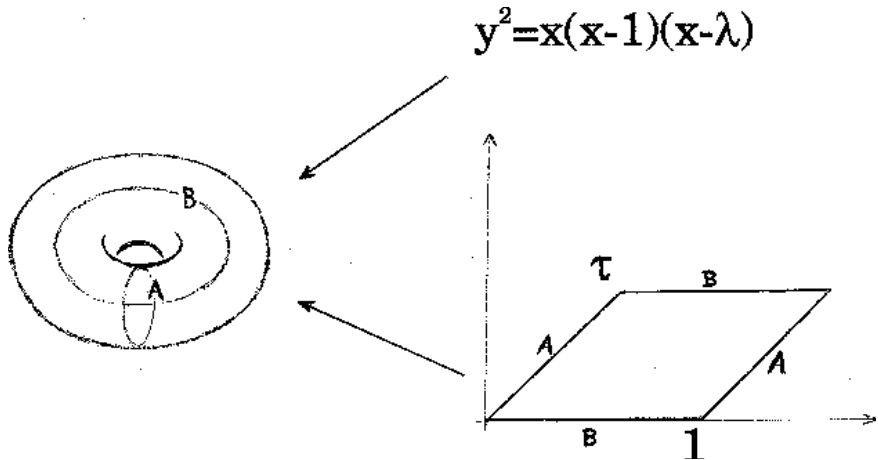,height=6cm}
\vspace{1cm}
\label{f.torus}
\caption{Elliptic curves or complex tori.}
\end{figure}

According to one of the many possible definitions 
(see e.g. \cite{elliptic}), an elliptic curve
is a complex curve of genus one (see fig. 2). There are two
alternative descriptions of these objects. 

The first one, very common in the physics literature dealing with CFT
and string theory, is the description  of elliptic curves as
complex tori $\C/(\Z+\Z\tau)$ parametrized by $\tau\in\C$ in the 
upper half-plane. Tori are obtained by identifying opposite edges in 
the parallelogram bounded by $0$, $1$, $\tau$ and $1+\tau$.

Obviously for some values of the parameters, say $\tau_1$ and
$\tau_2$, we may obtain indistinguishable tori. For example if we cut
through a non-contractible loop, rotate one edge through $2\pi$ and
glue back the edges, we obtain an equivalent torus. Such an operation
is called a Dehn twist. The notion of isomorphic tori makes this
more precise.

Two complex curves are isomorphic (i.e. can be considered as
indistinguishable) when there is a one to one holomorphic mapping
between them. It is natural to look for some parameter which
is identical for isomorphic tori, but enables us to distinguish
between distinct ones. Such a parameter is called a modulus. 
In fact one can associate with each elliptic curve a
complex number --- its $j$-invariant, which possesses precisely those
properties. Two elliptic curves are isomorphic {\em if and only if}
their $j$-invariants coincide. In this description 
the $j$-invariant is a well known transcendental function of $\tau$.
Moreover,  the symmetry which leaves $j$ invariant corresponds in this
description to modular invariance in the $\tau$-plane i.e.
\eq
j(\tau)=j(\tau')\;\;\Longleftrightarrow 
\tau'=\left(\f{a\tau+b}{c\tau+d}\right)
\quad \hbox{ for }
\left(\begin{array}{cc} 
a&b\\ 
c&d
\end{array}\right) \in\hbox{\slz} 
\eqx 
The geometrical meaning of this symmetry is as follows: all
modular transformations are generated by `Dehn twists'
(see e.g. \cite{Thiesen})
--- these are isomorphisms obtained by cutting the torus along one
loop, then twisting one edge through $2\pi$ and gluing it back.

At this point we see that if we could express all the operators in
our theory in terms of the $j$-invariant, we could reformulate
everything in terms of $\tau$ and obtain a modular invariant theory.

The reason why this might be interesting is that modular invariance
has proven to be a very strong constraint in QFT, allowing for
example for the classification of the partition functions of minimal models 
in CFT. Recently another application which could be relevant
in our case was the work of S-T. Yau and B.H.Lian on solution of
$3^{rd}$ order Fuchsian ordinary differential equations in terms of
modular forms \cite{Yau}.

Before we make the connection with the odderon, consider functions of
the form $\sqrt{j(\tau)-c}$. It turns out that the only possible
value of the constant $c$ for which this function is globally well
defined is 1728. Such an expression can be seen as parametrising tori
with a `sign'. We treat two tori as equivalent if one can be obtained
from the other by an {\em even} number of Dehn twists. Using the
above mentioned correspondence between Dehn twists and modular
transformations one can find that the invariance group is now an
infinite group generated by the transformations $\tau\ra\tau+2$ and
$\tau\ra -1/\tau$ . This group is the unique normal subgroup of
\slz of index 2 and is denoted by $\Gamma^2$.

\subsection{Modular invariance of the odderon}

To apply the preceding concepts to the odderon, we must use an
alternative but equivalent description of elliptic curves. This is
the Weierstrass parameterization which labels each elliptic curve by a
complex number $\lm\in \C$. The curve given by $\lm$ is given by the
equation
\eq
y^2=x(x-1)(x-\lm)
\eqx
where $x$ and $y$ are complex coordinates. This is also a complex
torus but presented in a different way. In fact the link between
those descriptions is given by the correspondence \cite{gm2}:
\eq
\lm(\tau)=\left(\f{\Theta_2(0;\tau)}{\Theta_3(0;\tau)}\right)^4
\eqx
where $\Theta_2(0;\tau)$ and $\Theta_3(0;\tau)$ are the Jacobi theta 
functions.

The $j$-invariant considered earlier can be expressed in terms of the
parameter $\lm$ labeling the elliptic curves.
It is now given by the formula:
\eq
j=2^8\cdot \f{(\lm^2-\lm+1)^3}{\lm^2(\lm-1)^2}
\eqx
Note that this expression is identical to the Vandermonde determinant
considered before (\ref{e.vander}), which is (up to the square root) 
the correct physical variable of the odderon. We see that our
physical variables (\ref{e.sqrt}) and (\ref{e.vander}) are just
conformal invariants of elliptic curves, and
using the more common description in terms of the $\tau$ parameter we
obtain a modular invariant theory with respect to the group
$\Gamma^2$. The benefit is that we may use an altogether different
set of tools which has proven to be very useful in a variety of
applications.

%Using the correspondence between $\lm$ and $\tau$ one  can reexpress
%the Hamiltonian and the integral of motion in terms of $\tau$.

\section{Conclusions}

In this talk we have presented the derivation of the modular
invariance of the odderon. We have shown that the global \slc invariance
and cyclic symmetry lead to the introduction of certain natural 
variables in terms of which the theory can be formulated. Furthermore
these quantities can be interpreted as conformal invariants of
elliptic curves with a `sign' --- the parity of the number of `Dehn'
twists. This at once leads to the modular symmetry of the theory.

Since the modular symmetry is a very strong constraint in QFT, we
hope that one can obtain some progress in solving the odderon,
especially as this symmetry opens up a new set of tools to attack the
problem. Furthermore it is very intriguing that such geometrical 
interpretation of the odderon symmetries exists. 
It is challenging to exploit this fact to obtain some relation with
an effective string theory.

\subsection*{Acknowledgments}
I would like to express my gratitude to Dr. Maciej A. Nowak for many
helpful suggestions.
I would like to thank Prof. Lev Lipatov for interesting discussion during
the conference.

\end{document}